\documentclass[12pt,preprint]{aastex}

\newcommand{\etal}{{\em et al.}}            

\shorttitle{{\em Chandra} Study of 3C\,403}
\shortauthors{Kraft \etal}

\begin{document}

\title{A {\em Chandra} Study of the Multi-Component X-ray Emission from the X-shaped Radio Galaxy 3C\,403}
\author{R. P. Kraft}
\affil{Harvard/Smithsonian Center for Astrophysics, 60 Garden St., MS-67, Cambridge, MA 02138}
\author{M. J. Hardcastle}
\affil{University of Hertfordshire, School of Physics, Astronomy, and Mathematics, Hatfield AL10 9AB, UK}
\author{D. M. Worrall}
\affil{University of Bristol, Department of Physics, Tyndall Ave., Bristol BS8 ITL, UK}
\author{S. S. Murray}
\affil{Harvard/Smithsonian Center for Astrophysics, 60 Garden St., MS-2, Cambridge, MA 02138}

\begin{abstract}

We present results from a 49.4 ks {\em Chandra}/ACIS-S observation of the 
nearby ($z$=0.059) X-shaped FRII radio galaxy 3C~403. This is the first
{\em Chandra} observation of an X-shaped radio galaxy, and one of
the goals of this pioneering study is to determine the
relationship between the X-ray emitting gas and the X-shaped
radio morphology. We find that the X-ray isophotes of the
hot gas within $\sim$3.5$''$ of the central galaxy are
highly elliptical (eccentricity$\sim$0.57) and co-aligned with the elliptical
optical isophotes. This supports the hypothesis that
X-shaped radio sources are created by propagation of jets
through asymmetric density distributions.  Within large uncertainties,
there is no evidence that the lobes or wings are overpressurized relative
to the ISM, supporting the scenario in which the wings are the result of
strong backflow of material from the jet head and subsequent buoyant evolution.
We have detected X-ray emission from several of the radio knots to the E of
the active nucleus, and diffuse emission from the radio lobe
to the W. The X-ray emission from the eastern knots cannot
be explained by an inverse Compton model unless they are far from
equipartition. Using archival HST data, optical emission is
detected from two knots, and the radio/optical/X-ray
spectra are well fitted by simple synchrotron models. 
This is one of the strongest examples to date of X-ray synchrotron emission
from multiple knots in the jet of an FR II radio galaxy.
X-ray emission is also detected from the radio wings at a flux
consistent with inverse Compton scattering of CMB photons from
relativistic electrons if the wings are near equipartition.
The nuclear spectrum is well described by a multi-component
model that includes a heavily absorbed power law
($N_H\sim$4$\times$10$^{23}$ cm$^{-2}$) and a bright
(EW$\sim$244 eV), broadened Fe line. A second, less absorbed, power-law
component, likely to represent unresolved emission from a pc-scale jet, is also required.
\end{abstract}


\section{Introduction}

Approximately 10\% of the FRII radio galaxies in the 3CRR catalog are
classified as `X-shaped' or 'winged' because of their unusual radio
morphologies \citep{lea84}.  These radio galaxies have two pairs of misaligned radio
lobes of approximately equal linear extent.  They also tend to have
low radio luminosities, typically lying near the FRI/FRII division.
There are two general hypotheses about the nature and origin of these structures.
On the one hand, it has been argued that the X-shaped morphology is
essentially a hydrodynamic phenomenon which is the result of supersonic
or buoyant flow of radio plasma in an asymmetric gas distribution.
\citet{lea84} and \citet{wor95} argued that the X-shaped radio
morphology is a result of the result of strong backflow of material behind the terminal hot spots of 
radio galaxy jets and subsequent buoyant evolution of the 'wings'.
\citet{cap02} hypothesized
that the X-shaped morphology is the direct result of the supersonic expansion/inflation
of the lobe into an elliptical atmosphere, and that all radio galaxies in such environments
should exhibit this phenomenon.
On the other hand, it has been argued that this
structure results from precession or reorientation of the jet axis, possibly as
a result of a galaxy interaction and/or merger event of two
supermassive black holes \citep{eke78,gop03,den02}.
In this scenario, the 'wings' represent the remnant lobe of an earlier epoch
of nuclear activity, and only the primary 'jet/lobe' is currently being powered
by the nuclear outflow.
This latter scenario has generated particular interest recently
because it implies that these galaxies may be strong sources of gravitational radiation \citep{mer02}.

The sensitivity of Chandra and XMM-Newton has been a crucial
factor in making progress in the study of the gaseous environments of
the radio galaxies and better understanding issues related to the
dynamics of the radio structures. 
In addition to emission from the thermal coronae,
{\em Chandra} and XMM-Newton observations of radio galaxies
have detected dozens of examples of
non-thermal X-ray emission from their jets and hotspots.
The standard paradigm suggests that the X-ray emission observed from
these features has different origins depending on the power of the
radio galaxy.  The X-ray emission from the knots of jets of FR I
radio galaxies are generally attributed to synchrotron emission from
a population of ultra-relativistic electrons \citep{wor01}. 
This conclusion is supported by several pieces of evidence including
their broad-band spectra.
Well studied examples include Centaurus A \citep{kra02,hard03}, 
M87 \citep{mar02}, and 3C 66B \citep{hard01}.  On the other hand, inverse-Compton 
scattering of seed photons by relativistic electrons is generally believed 
to be responsible for the X-ray emission from jets \citep{tav00,sam01} and
hotspots \citep{hard04} of the more powerful FRII sources.
Well known hotspot examples include Cygnus A \citep{car94} and 3C 295 \citep{har00}.
To date, only two examples of X-ray emission from FR II hotspots have
been attributed to synchrotron emission, 3C 390.3 \citep{har98} and
Pictor A \citep{wil01}, although \citet{hard04} argue that the X-ray emission from a large
number of low power hotspots is at least partly synchrotron in origin.

3C\,403 is the nearest X-shaped radio galaxy in the 3CRR catalog \citep{lai83}, and is
an ideal target to study the X-shaped phenomenon.
In this paper we present results from a {\em Chandra}
observation of this radio galaxy.  
We use archival VLA and HST observations
to make a multi-frequency study of the various components.
We have resolved X-ray emission from four distinct components:  the bright active nucleus,
the hot ISM, several compact radio components, and diffuse X-ray emission coincident
with the radio lobes.  
This is the first high-resolution X-ray observation of an X-shaped radio galaxy
and the primary goal of this observation was to study the ambient ISM confining the radio emitting components.
In addition, the non-thermal X-ray emission that we detect from several of the compact radio components
allows us to draw strong conclusions regarding their emission mechanisms.

This paper is organized as
follows.  Section two contains a brief description of the data, an
observation log, and a summary of our analysis.
In section three, we discuss the emission from the hot ISM and
nucleus, and we describe how we have separated these components spatially and spectrally.  
In section four, we discuss the non-thermal X-ray emission associated with the
radio hotspots and lobes.  The nature of X-shaped radio galaxies
and the implications of our results are discussed in section five.
The final section contains a brief summary and conclusion.
Throughout this paper, we assume the WMAP
cosmology ($H_0$=71, $\Omega_M$=0.27, $\Omega_\Lambda$=0.73), so that the
measured redshift of the host galaxy ($z=0.059$) corresponds to a luminosity
distance of 260.6 Mpc, and 1$''$ = 1.127 kpc.

\section{Observations}

The radio galaxy 3C\,403 was observed for 49470 s with the {\em Chandra}/ACIS-S
instrument (OBSID 2968) in faint mode on December 12, 2002.  The nucleus of the galaxy was located at
the best focus.  The lightcurve of events in the 5.0-10.0 keV 
bandpass on the entire S3 chip, excluding the nucleus and any other point sources
visible by eye, was created using 259 s bins and examined for periods of flaring background. 
Intervals where the rate was more than 3$\sigma$ above the mean rate were removed.
The mean rate with periods of flaring removed is consistent with the nominal quiescent
ACIS-S background listed in the Proposer's Observation Guide.
A total of 4600 s of data was excluded, leaving 44870 s of good time.  Bad pixels,
hot columns, and columns along node boundaries were also removed.  
This galaxy is located at relatively low Galactic latitude ($\ell=42.263$, $b=-12.314$),
so the nominal blank sky background files generated from deep, high Galactic latitude
observations are not appropriate.
Fortunately the radio galaxy covers only a small region of the S3 chip,
so that local background for spectral
analysis was estimated from a large region on the S3 chip north of the source.
Absorption by foreground gas in our galaxy ($N_H$=1.54$\times$10$^{21}$ cm$^{-2}$) was
included in all spectral fits.

\begin{figure}
\plotone{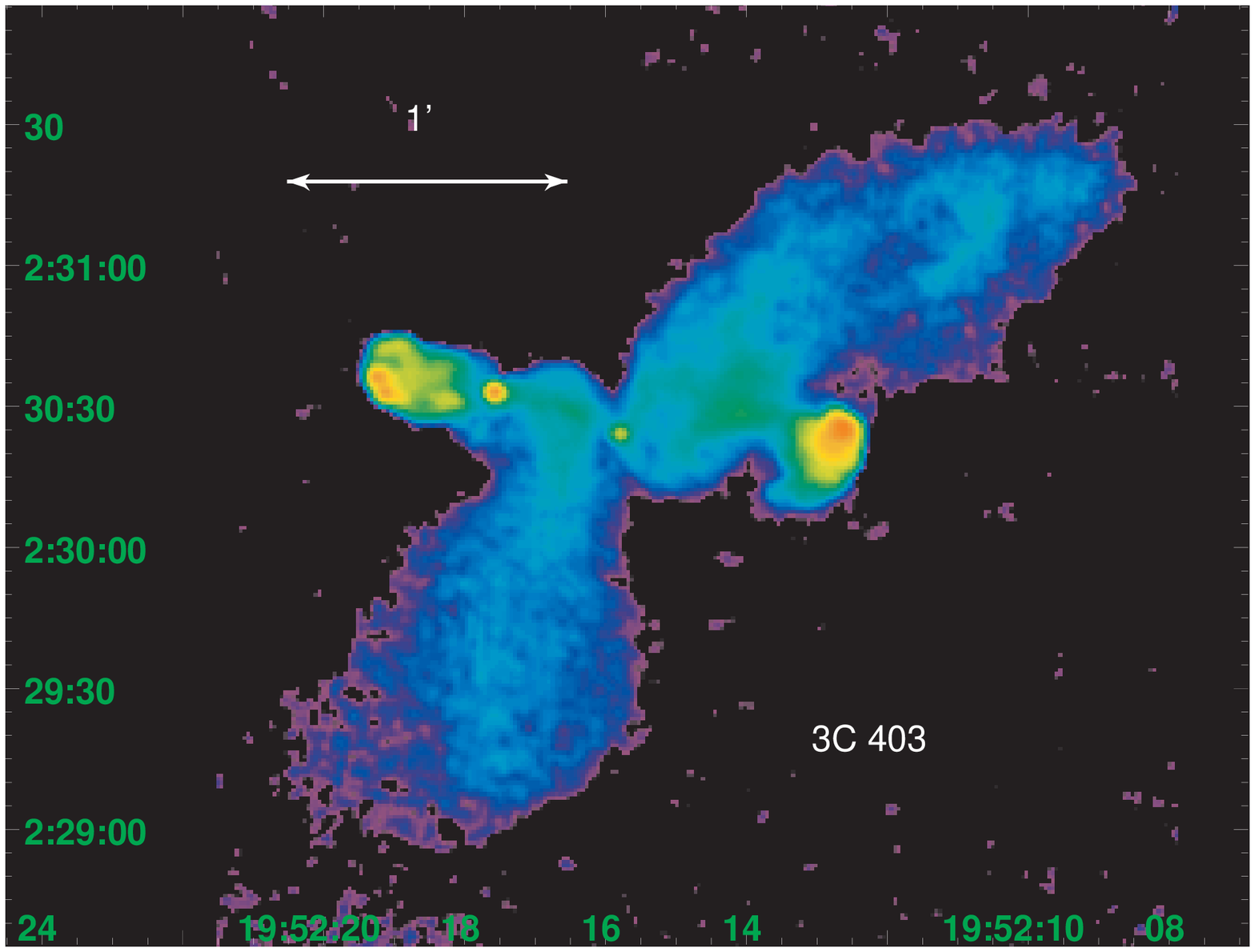}
\caption{8.4 GHz radio map (0.75$''$$\times$0.75$''$ beam (FWHM))
of 3C\,403 \citep{bla92}.  The primary radio lobes and the secondary
radio 'wings' lie along the E/W and NW/SE axes, respectively.}\label{radioimg}
\end{figure}

We have combined data from our {\em Chandra}/ACIS-S observation with archival 
VLA observations (1.48 and 8.4 GHz) and an HST observation
to make a multi-frequency study of various components.
Detailed discussions of the 1.48 and 8.4 GHz VLA data have been presented in \citet{den02} and
\citet{bla92}, respectively.  We will
use the nomenclature of the \citet{bla92} throughout this paper
when referring to specific knots or radio features.
An 8.4 GHz radio map of this radio galaxy is shown in Figure~\ref{radioimg}.
The primary lobes can be seen to the E and W of the central nucleus.  The lower
surface brightness `wings' that give this object its X-shaped morphology extend
to the N and S.  Several hotspots can be seen in the eastern lobe, and perhaps
a hint of a jet between the nucleus and knot F6.
The X-ray and radio positions of the active nucleus agree to $\sim$0.5$''$.
This radio galaxy was also observed for 280 s with the HST/WFPC2 instrument
in 1995 as part of the HST 3CR snapshot survey using the
F702W filter (pivot wavelength of 6919 \AA) \citep{mar99}.
We obtained the reprocessed data from the HST archive and used
the IRAF synphot package to apply photometric calibrations.
The fluxes were reddening-corrected using the dust maps of \citet{sch98}.

\begin{figure}
\plotone{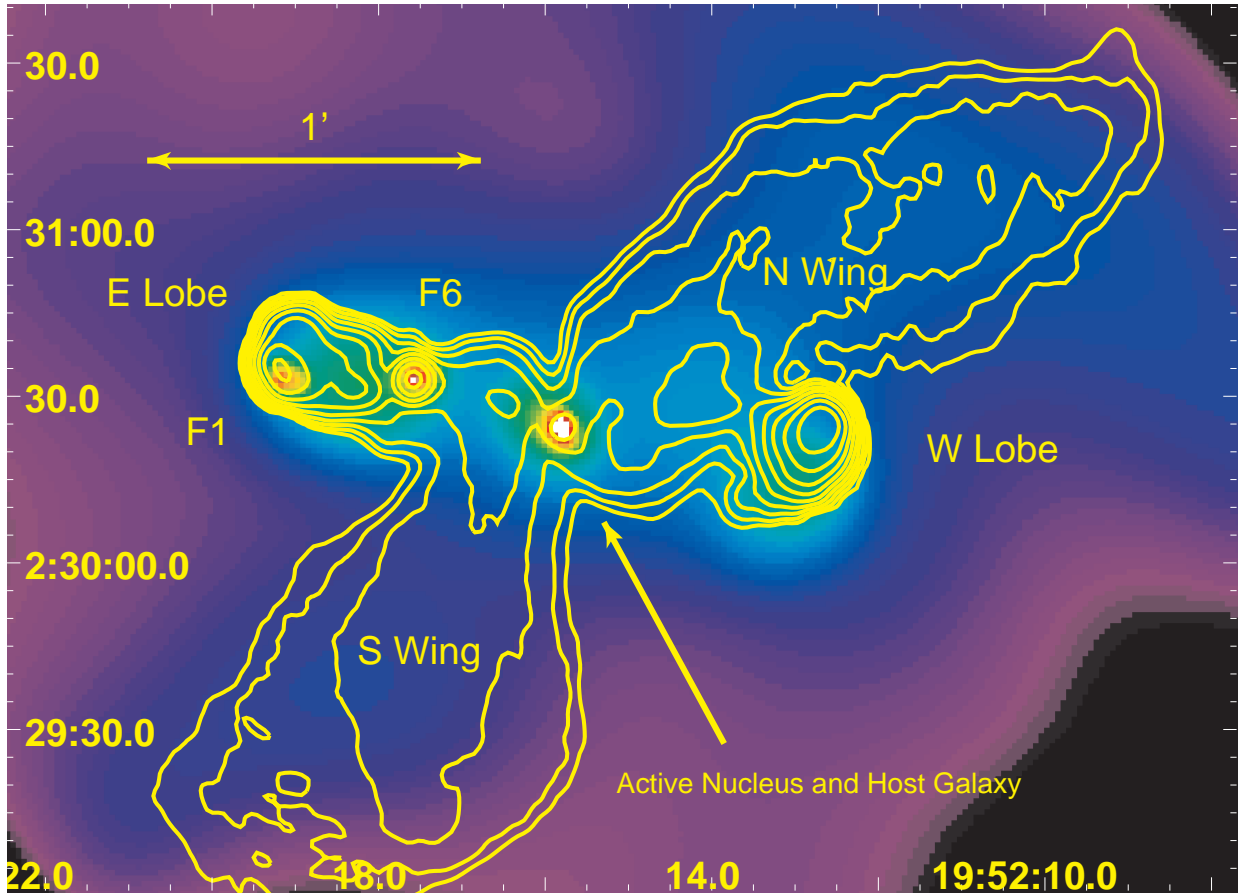}
\caption{Adaptively smoothed, exposure corrected, background subtracted
{\em Chandra}/ACIS-S image of 3C\,403 in the 0.5-2.0 keV band
with 8.4 GHz radio contours overlaid.  We
have detected emission from the active nucleus, the hot ISM, several of the
compact radio components, and diffuse emission from the lobes and wings.}\label{xradovl}
\end{figure}

An adaptively smoothed, exposure corrected, background subtracted X-ray image in
the 0.5-2.0 keV bandpass with 8.4 GHz radio contours overlaid is shown in
Figure~\ref{xradovl}. 
X-ray emission is detected from at least four distinct components: the active
nucleus, the hot gas of the host galaxy, several of the radio hot spots, and diffuse
emission from the lobes and the wings.

\section{Hot ISM and Nucleus}

\begin{figure}
\plotone{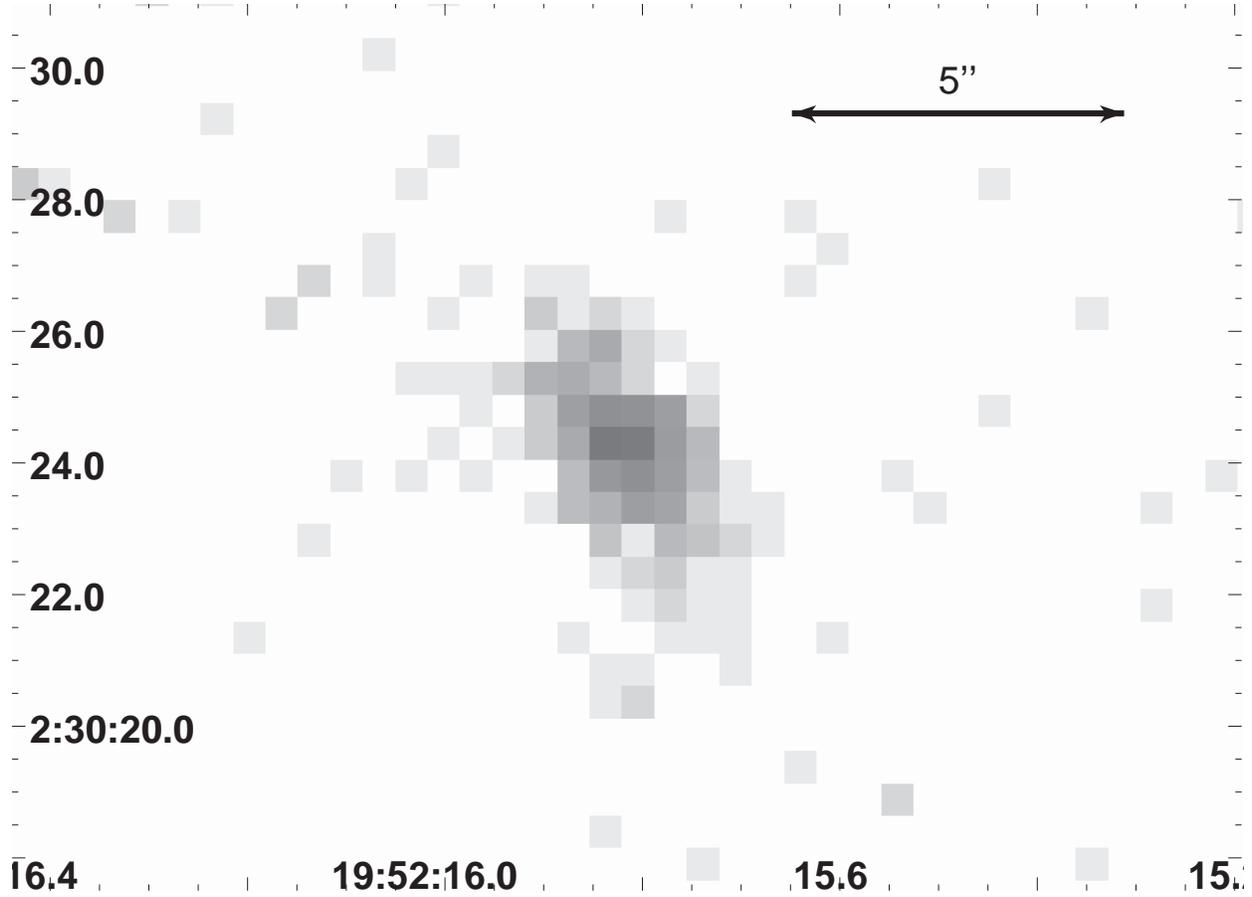}
\caption{Raw {\em Chandra}/ACIS-S image of the central 20$''$$\times$15$''$ region of
3C\,403.  The unresolved point source is at the center, and the emission of the
hot ISM extends to the NE and SW of the nucleus.  The brightest pixel
contains 63 events, and there are 515 total in the image.}\label{central}
\end{figure}

A raw X-ray image in the 0.5-2.0 keV bandpass
of the central region (20$''\times$14$''$), including the hot ISM and the active nucleus,
is shown in Figure~\ref{central}.
The point source at the center is the AGN, and the extensions to the NE and SW are the
hot ISM.  The morphology of the ISM is highly elliptic, and co-aligned with, but more
eccentric than, the optical isophotes
of the stellar light distribution \citep{cap02}.
We wish to separate the emission of the hot ISM from that of the active nucleus
in order to study each component separately.  Unfortunately, these components
are not well separated spatially.
The major and minor axes of the ISM shown in Figure~\ref{central}
are approximately 7.5$''$ and 3.0$''$ in length, respectively.
The 90\% encircled energy radius for a point source on-axis at 1.49 keV is $\sim$0.9$''$.
We have used images in four X-ray bands to guide our spectral analysis in order to
separate the emission from the two components.
An examination of images in the 0.5-0.7, 0.7-1.0, 1.0-1.5, and 1.5-2.0 keV bands showed that
above 1 keV, most of the emission is unresolved and is probably due to the active nucleus.
Below 0.7 keV, the contribution from the central point source is
small, and most or all of the emission is from the ISM.
 
\begin{figure}
\plotone{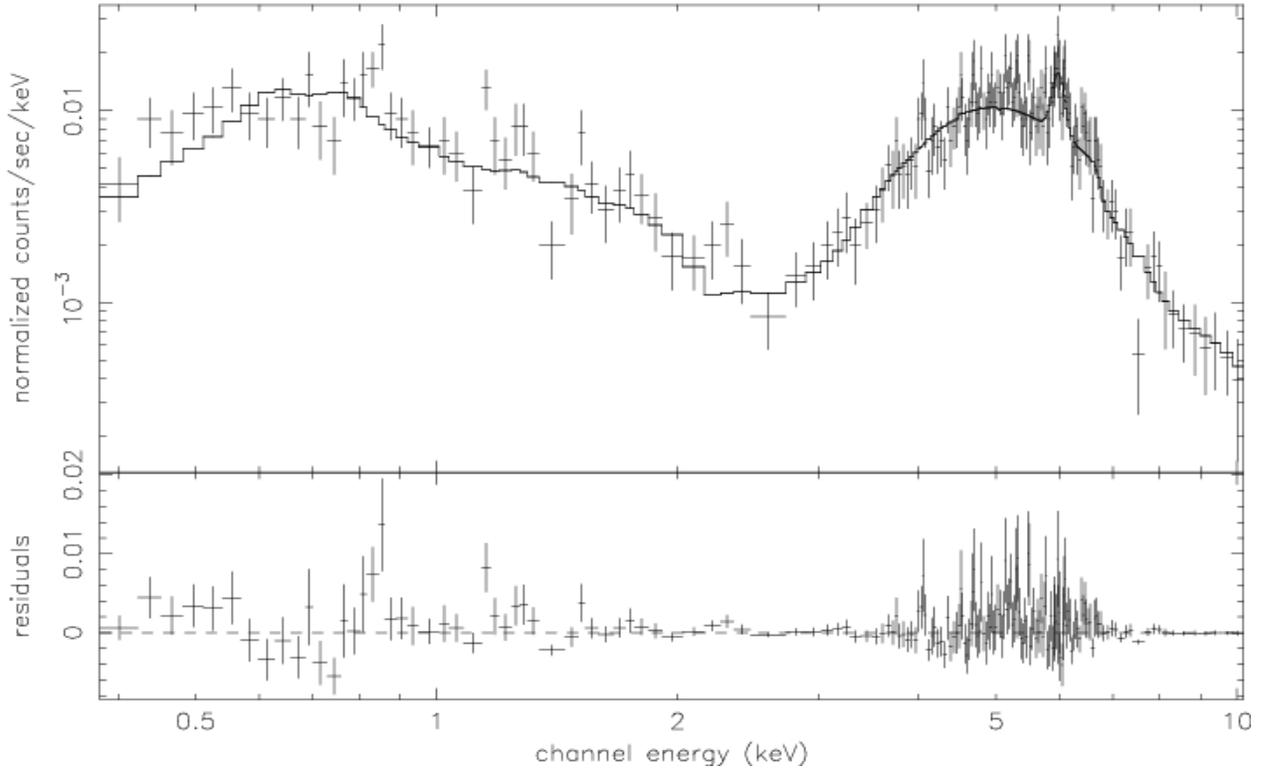}
\caption{X-ray spectrum of the central 7.5$''$ radius region with best-fit
model overlaid.  The residuals are plotted underneath.
The best-fit parameters and uncertainties are summarized
in Table~\ref{specparms}.  This spectrum includes emission from the unresolved core
and the hot ISM.}\label{spectrum}
\end{figure}

We extracted the spectrum, shown in Figure~\ref{spectrum}, from a 7.5$''$ radius circular
region centered on
the nucleus.  The data have been binned to a minimum
of 10 counts per bin, and the background was determined from a large region on the S3
chip approximately 4$'$ to the N of the nucleus, well removed from any detectable emission from
the galaxy.  The spectrum consists of two peaks: one centered at $\sim$0.7 keV
with a high energy tail, and a second that peaks at around 5 keV.
The count rate in the 0.25-10.0 keV band was 4.7$\times$10$^{-2}$
cts s$^{-1}$.  The majority of the counts in this spectrum are from the unresolved
active nucleus.
This suggests that this spectrum moderately suffers from the effects of pile-up.
Based on Figure 6.19 of the {\em Chandra} Proposer's Observatory Guide (POG), we estimate
a pile-up fraction of $\sim$5\%.
We note that we fitted the data using coarser binning (25 and 50 counts per bin) to
ensure the applicability of Gaussian statistics and find statistically identical parameter
values and uncertainties.

We initially modeled the spectrum with a three-component model including a heavily absorbed
($N_H>$10$^{23}$ cm$^{-2}$) power law,
an Fe line with variable centroid and width, and an APEC model.  The effect of the pile-up was
modeled by using the XSPEC function `pileup' \citep{dav01}, and Galactic absorption
($N_H$=1.54$\times$10$^{21}$ cm$^{-2}$) was included in all fits.
For simplicity, the elemental abundance of the APEC model was fixed at 1.0 times Solar,
typical of the central regions of early-type galaxies.
If allowed to vary freely, the elemental abundance was poorly constrained.
At the relatively low best-fit temperature
($\sim$0.3 keV), the emission is line dominated, and
the elemental abundance can be traded against the normalization.
The radiative cooling function, $\Lambda (T,Z)$, scales nearly linearly with abundance at 
these temperatures and abundances.
None of our conclusions below are sensitive to this choice, although
a larger elemental abundance would
reduce our estimates of the gas density below ($n_H \sim Z^{-1/2}$).

\begin{figure}
\plotone{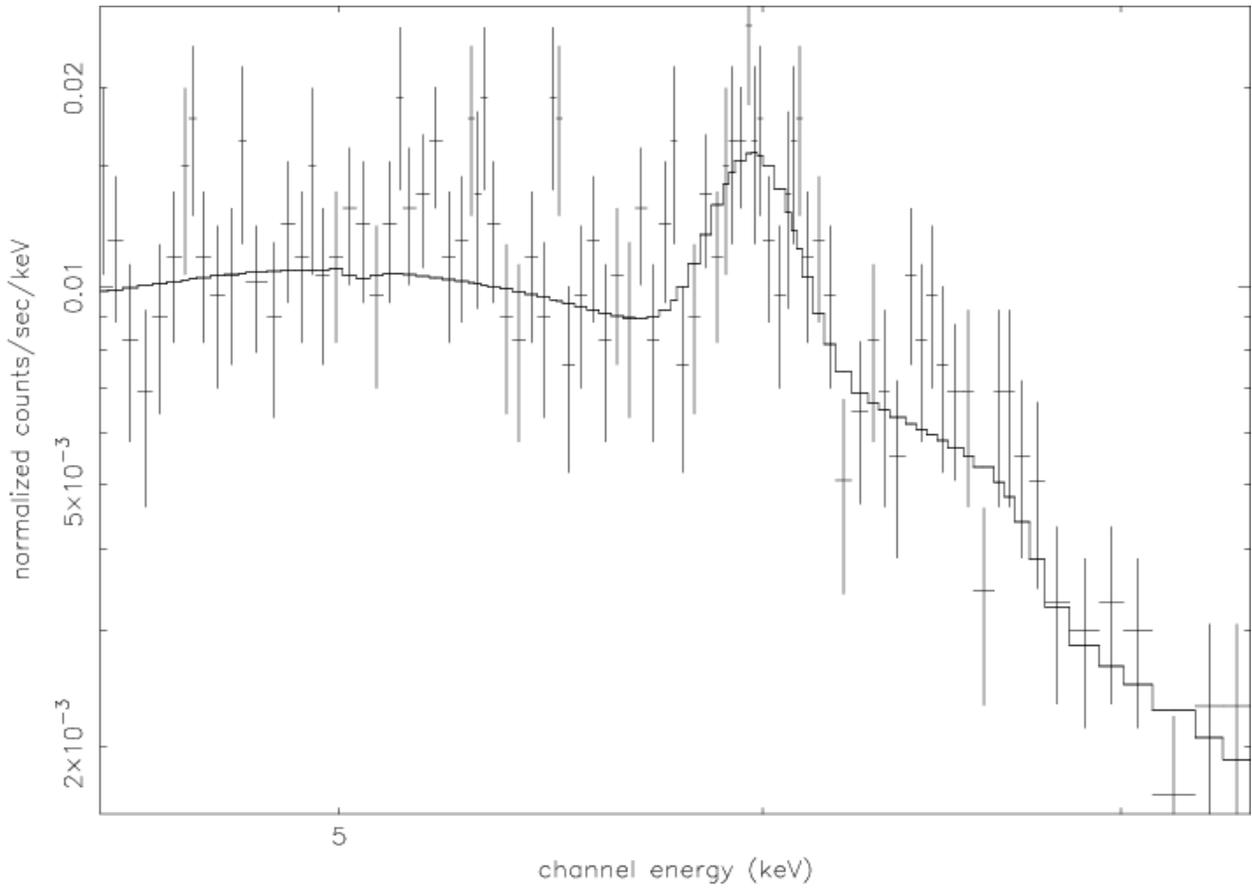}
\caption{Data and best-fit model X-ray spectrum of the central 7.5$''$ radius region 
around the Fe K line.}\label{fespectrum}
\end{figure}

In this scenario, the power law represents emission from the AGN, and the APEC model the
emission from the hot ISM.  We could not obtain an acceptable fit with this
three component model as there were
large residuals between 1 and 2 keV.
Inspection of images in the 1-2 keV band suggests that most of the emission in this band is
is due to the unresolved AGN.
We therefore added a second, less absorbed power law to the model and were able to obtain an
acceptable fit.  An examination of the images in several bands (described above)
suggests that this second power law is heavily absorbed.
We fixed the absorption and spectral index of this second power law
to 4$\times$10$^{21}$ cm$^{-2}$ and 2.0 respectively to minimize the number of free
parameters; only the normalization of this component was allowed to vary freely.
The choice of a spectral index that is steeper than that
of the primary absorbed power law and the choice of the value 
of $N_H$ constrain this component to contribute little above and below the 1-2 keV band.
The best-fit model has been overlaid onto Figure~\ref{spectrum}.
The data and model around the Fe K line is more clearly shown in Figure~\ref{fespectrum}.
The best-fit values of the parameters and 90\% confidence intervals are summarized in
Table~\ref{specparms}.

\subsection{ISM}

\begin{figure}
\plotone{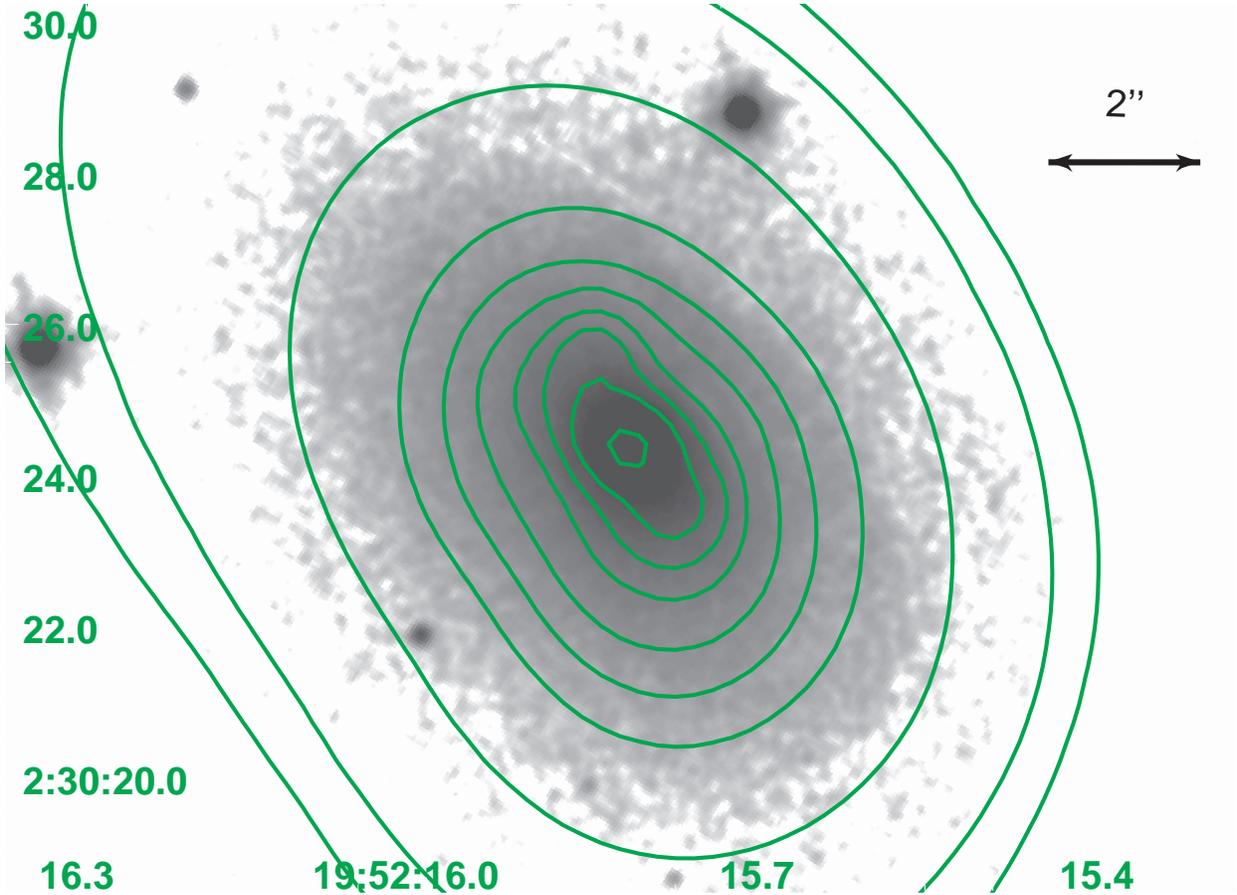}
\caption{Contours from an adaptively smoothed X-ray image in the 0.3-1.0 keV bandpass
overlaid onto an HST/WFPC2 image of the host galaxy of 3C\,403.  The ellipticity and
angle of rotation of the optical and X-ray isophotes are consistent.}\label{xoptovl}
\end{figure}

The morphology of the X-ray emission of the central region of
the hot ISM is highly elliptical and co-aligned with the optical
isophotes of the host galaxy.  We have overplotted the contours from an adaptively smoothed
X-ray image in the 0.5-2.0 keV band onto an HST/WFPC2 image of the host galaxy as shown
in Figure~\ref{xoptovl}.  The position angle of the diffuse emission is not
aligned with the primary E/W radio components, and is almost perpendicular to
the N and S radio wings.
The spatial relationship between the lobes, the wings, and the ISM, at least near
the nucleus, is qualitatively what we would expect in the buoyancy/backflow model.
Based on the spectral analysis above,
the unabsorbed X-ray luminosity of the ISM within 8.5 kpc of the nucleus
is 2.9$\times$10$^{41}$ ergs s$^{-1}$ in the 0.25-10.0 keV bandpass.
This X-ray luminosity is at the high end of the $L_X$ versus $L_B$ relationship derived
from observations of large samples of elliptical galaxies in \citet{can87} and \cite{ewan01}. 
The temperature of the gas in this central region is $\sim$0.3 keV, rather low
for a massive elliptical galaxy, perhaps indicative of a cooling flow.

We fitted an elliptical $\beta$-model plus constant background to an X-ray image of the central
5$''$ of the galaxy in the 0.3-1.0 keV bandpass using the Sherpa software
package in order to characterize quantitatively the
morphology of this emission.
As demonstrated above, the contribution from the nucleus in this band is small.
The free parameters of this model were $\beta$, the
scale parameter $r_0$, the normalization, the ellipticity, $\epsilon$, the
position angle of the major axis of the ellipse, $\theta$, and the background, $B$.  The telescope
PSF was computed using the CIAO program {\em mkpsf} to remove the instrumental
response.  Both $\epsilon$ and $\theta$ were well constrained, with $\epsilon$=0.57$\pm$0.04
and $\theta$=44$^\circ\pm$2$^\circ$ (measured East of North).  The uncertainties
are at 90\% confidence.
Based on the spectral analysis described above, the 0.3-1.0 keV band contains little
emission from the active nucleus.
The parameters $\beta$ and $r_0$ were poorly constrained, but $\beta$$>$0.5
and $r_0$$<$3$''$ ($<$3.5 kpc).  Using nominal values of $\beta$=0.55 and $r_0$=0.6$''$, the
central proton density, $n_0$, is $\sim$1.14 cm$^{-3}$ for the best-fit spectrum
described above. 
The values of $r_0$ and $n_0$ are somewhat smaller and larger, respectively, than
typically observed in elliptical galaxies, although it is
not uncommon to find small scale, dense cores
in the ISM at the centers of radio galaxies \citep{hard02}.
The cooling time of this material is very short, $\sim$10$^7$ years, because of the
high density and relatively low temperature.
The cooling time is insensitive to the adopted abundance, since
a metal abundance higher than we have assumed (1.0 times the Solar value)
would decrease the central density but increase the cooling time (the cooling time
scales roughly as Z$^{-1/2}$).

On the kpc scale, the X-ray morphology, and presumably the density and pressure profiles,
are highly elliptical ($\epsilon$=0.57).  In contrast, the ellipticity of the stellar light profile
is only 0.25 at 6.9$''$ (the effective radius of the de Vaucouleurs profile for the host of 3C\,403)
from the nucleus \citep{gov00}.
We have detected X-ray emission to only half this distance along the semi-major axis, but an
examination of the central 3$''$ of the HST images clearly shows that
the stellar distribution has a roughly constant ellipticity ($\sim$0.25) down to about
1$''$, where the distribution becomes circular.
The distribution of the gas is therefore more elliptical than that of stars.
In the central 10-15 kpc of a elliptical galaxies, the stellar mass typically 
dominates the dark matter
so the gravitational potential should be determined by the distribution of the former.  If the gas
is in hydrostatic equilibrium with the gravitating matter, we expect that the ellipticity of the
gas and that of the gravitating matter should be similar, 
but in fact this is not the case in 3C\,403.

We suggest two possible explanations for this.  The first, and most likely in our
view, is that the gas
is not in hydrostatic equilibrium.  The cooling time of this gas is less than 10$^7$ years,
roughly the same age as the radio lobes.
It is plausible that the gas is cooling and falling in toward the center.
The inflation of the radio lobe may have disrupted the central region as well.
If the gas is moving, it may well have a different morphology than the gravitating matter.
As an alternative, it is possible that the emission which we have been
attributing to the ISM is, in fact, due to a partially resolved kiloparsec scale jet.
We consider this unlikely for two reasons.  First, there is no evidence for
any radio counterpart to this emission.  The X-ray to radio spectrum would have to
be unusually flat.  Second, the axis of this hypothetical kpc
scale jet is not aligned with the larger scale, tens of kpc radio features, which
implies that the jet has been sharply bent.  

\begin{figure}
\plotone{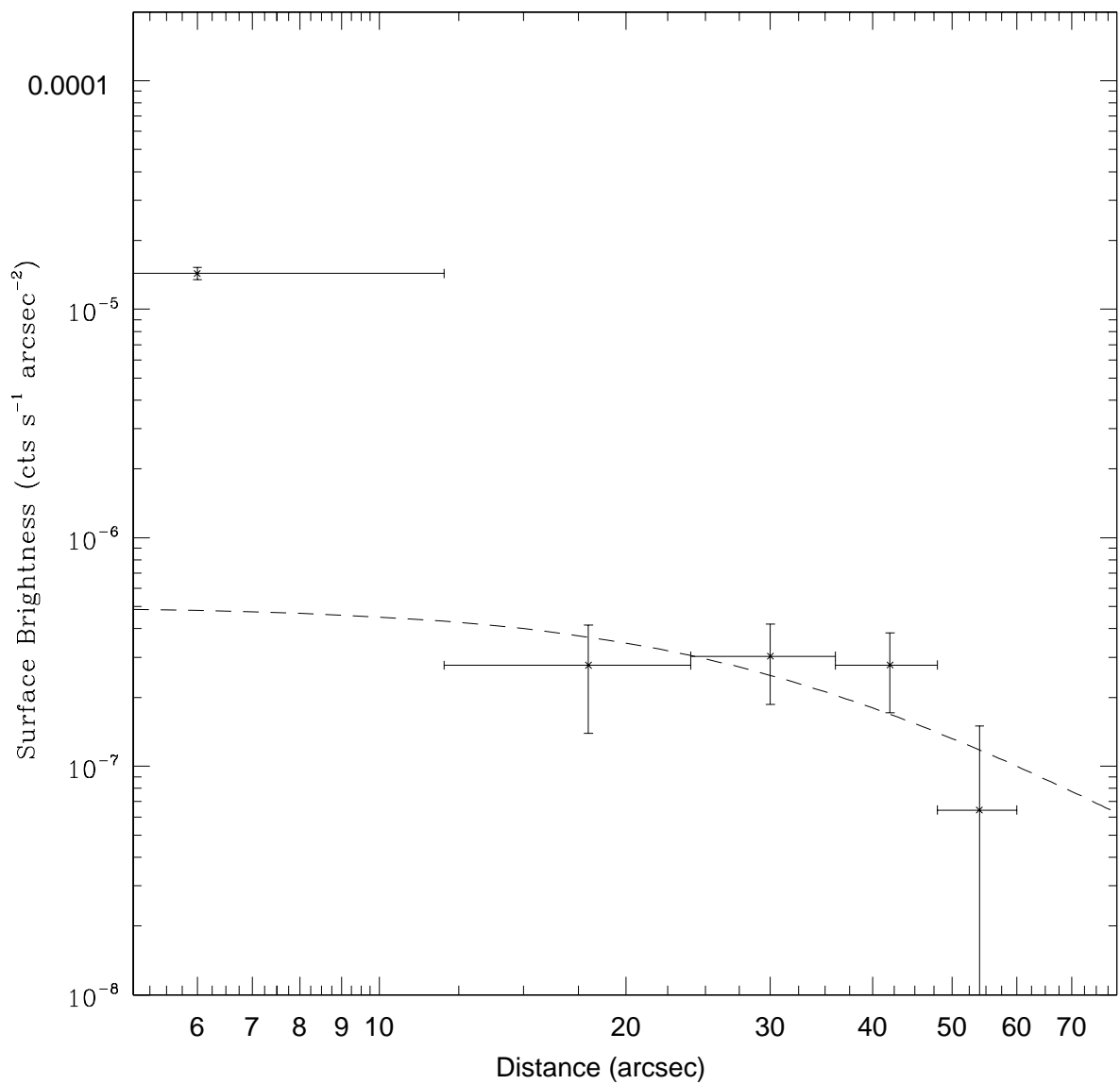}
\caption{Radial surface brightness profile (background subtracted)
in circular annuli of the larger
scale (tens of kpc) diffuse emission of 3C\,403 in the 0.3-1.0 keV band.
All the point sources, knots, and diffuse emission from the W radio lobe have
been removed.  The surface brightness of the background is
6$\times$10$^{-7}$ cts s$^{-1}$ arcsec$^{-2}$ and was estimated 
from a 1.5$'$ radius circular
region $\sim$4$'$ to the N of the nucleus.  The dashed curve represents the
best-fit $\beta$-profile excluding the first data point.}\label{sbprof}
\end{figure}

On larger (tens of kpc) scales, the data are not of sufficient quality for us to determine
whether the gas distribution remains elliptical at or beyond the boundaries of the wings and lobes.
We can, however, address the simpler question of whether the radio wings
are greatly overpressurized relative to the ambient medium.
There is statistically significant emission above background out to about 50$''$ from the
nucleus, although the profile of the emission is poorly constrained.  This could represent
emission associated with a group of galaxies for which the optical host of 3C\,403
is the dominant member.  
The surface brightness profile of the gas from the core to $\sim$50$''$
cannot be described by a single $\beta$-profile, however.
Given the small core radius and high density of the emission in the central region,
it is not reasonable simply to extrapolate this to larger radii to estimate the gas pressure
for a comparison with the pressure of the lobes and the wings.
For simplicity, and because it would be difficult to maintain a gas distribution to
large radii that is more elliptical than the stellar mass distribution, we assume
that the profile is spherically symmetric and parameterize the
data with a circular $\beta$-profile.  We note that
using an elliptical profile (e.g. with the same ellipticity that we found
for the smaller scale emission) would not significantly change our conclusions below. 
A plot of the surface brightness in circular annuli versus distance from
the nucleus is shown in Figure~\ref{sbprof}.
X-ray knots, hotspots, and background point sources have been excluded.

We measure a flux of 1.32$\times$10$^{-14}$ ergs cm$^{-2}$ s$^{-1}$ (unabsorbed)
in the 0.1-10.0 keV band in an annular region 50$''$ in radius from the nucleus, excluding
the central 10$''$.  The X-ray luminosity of this emission is $\sim$1.1$\times$
10$^{41}$ ergs s$^{-1}$ in the same band.
We have assumed a gas temperature of 0.7 keV, typical for groups, and an elemental abundance of
0.4 times Solar.  For an assumed spherical $\beta$-model profile with $\beta$=0.5 and
$r_0$=30$''$, the central proton density is $\sim$3.3$\times$10$^{-3}$ cm$^{-3}$.  
The gas pressure at the distance of the
center of the wings is $\sim$2$\times$10$^{-12}$ dyn cm$^{-2}$.
The equipartition pressure of the wings is $\sim$5$\times$10$^{-13}$ dyn cm$^{-2}$ (see
Section 4.2).  
If all of the X-ray emission from the wings is due to IC scattering of CMB photons, the pressure is
$\sim$1$\times$10$^{-12}$ dyn cm$^{-2}$ (see below).
Thus, the internal minimum pressure of the radio wings is the same order of magnitude
as the pressure of the ambient medium.  Our estimate of the gas pressure has large uncertainties, and
we have ignored the effects of projection which would place the lobes further out
into the halo and into lower density gas.  Given the uncertainties, it is reasonable
to conclude that the lobes are in approximate pressure equilibrium with the ISM.
There is no evidence to suggest that the
wings are greatly overpressurized relative to the ambient medium.

\subsection{Active Nucleus}

The spectrum of the nucleus is well described by a two power-law model plus an
Fe line.  The primary power-law is heavily absorbed ($N_H\sim$4$\times$10$^{23}$ cm$^{-2}$).
We attribute the primary power-law component to emission from material
near the inner part of the accretion close to the central supermassive black hole.
There is no evidence for a significant amount of dust in HST observations
of the host galaxy \citep{mar99}, so it is likely that this large
absorption column is due to cool material in close proximity
to the central black hole as well, perhaps a molecular torus.
This is consistent with previous optical emission line studies that have
shown that 3C\,403 is a narrow emission line galaxy (NELG) \citep{tad93,tad98}.
This suggests that we are observing the disk/torus nearly edge on, and that
the jets are advancing at an angle close to the plane of the sky. 
The broad line region, if it exists, must be buried behind the torus and unobservable.
Neither the photon index (1.7$\pm$0.15) nor the unabsorbed X-ray luminosity
(1.07$\times$10$^{44}$ ergs s$^{-1}$ in the 0.25-10 keV bandpass) are unusual for the nuclei
of radio galaxies.

The parameters of the second power-law component are poorly constrained, and
as described above, we have fixed the power-law index ($\alpha$=2)
and absorption ($N_H$=4$\times$10$^{21}$ cm$^{-2}$) of this component
for consistency with the imaging data.
We suggest that this second power-law component is X-ray
emission from an unresolved jet that is partially obscured, as observed with
greater sensitivity in Centaurus A \citep{eva04}.
The radio luminosity of the core is 6.5$\times$10$^{21}$ W Hz$^{-1}$ sr$^{-1}$
at 5 GHz.  The X-ray luminosity of the second power-law component
is 7.1$\times$10$^{15}$ W Hz$^{-1}$ sr$^{-1}$ at 1 keV.
The X-ray to radio flux ratio is consistent with that found in a sample of
low-power radio galaxies \citep{can99}, and supports the hypothesis
that this second power-law component is related to spatially unresolved
emission from a jet.
It is likely that the excess column density is due to absorption within a few
pc of the nucleus.
We speculate that this unresolved jet lies above the accretion disk, and our
line of sight to the jet passes through a region of lower column density
higher up in the torus.
There is no evidence of gas or dust on kpc scales, perhaps from a recent merger,
that could be responsible for this large column.
It is also possible that the hot ISM is responsible for this large column.
The column density of the hot gas within a distance $r_0$ (the central
core radius) is $N_H^{ISM}= r_0 \times (1-\epsilon)\times n_0\sim$
1.0$\times$10$^{21}$ cm$^{-2}$, remarkably close to the 4$\times$10$^{21}$ cm$^{-2}$
value chosen based on the multi-band images.  A careful calculation would require 
detailed knowledge of the ionization state of the gas, the distribution of gas along the line of
sight, and the elemental abundance, all of which are poorly constrained.
Alternatively, it is possible that this second power-law component is due either
to electron scattering from an ionized corona or a partial covering fraction
absorber, both of which are often invoked to explain the X-ray spectrum of
Seyfert galaxies \citep{tur97}.
The observed flux ratio of the first and second power-law components is consistent
with that observed in Seyfert galaxies.
The jet hypothesis is, however, more natural in our view as 3C 403 is an FRII
radio galaxy.  This hypothesis could be evaluated by a VLBI observation to search for
a pc-scale jet near the core.

A strong Fe line with EW=244$\pm$20 eV (90\% confidence) has also been detected.
The centroid of the line is 6.31$\pm$0.04 keV, and the width (r.m.s.) of the line is 80$\pm$50 eV.
The centroid of this line is consistent with fluorescent emission from cold
material.  The large EW of this line is consistent with a model where the line emission
originates from an absorbing region that surrounds the central supermassive
black hole \citep{lea93,miy96}.  The measured width of the Fe line corresponds to
a velocity (FWHM) of 900 km s$^{-1}$.  If the line broadening is the result of
fluorescence from cold gas in Keplerian motion around a 10$^9$ M$_\odot$ black hole, 
it must lie $\sim$5 pc from the black hole.  It is reasonable to attribute the large
column and fluorescent Fe line to a molecular torus that surrounds the central
supermassive black hole.

\section{Extended radio-related X-ray emission}

\begin{figure}
\plotone{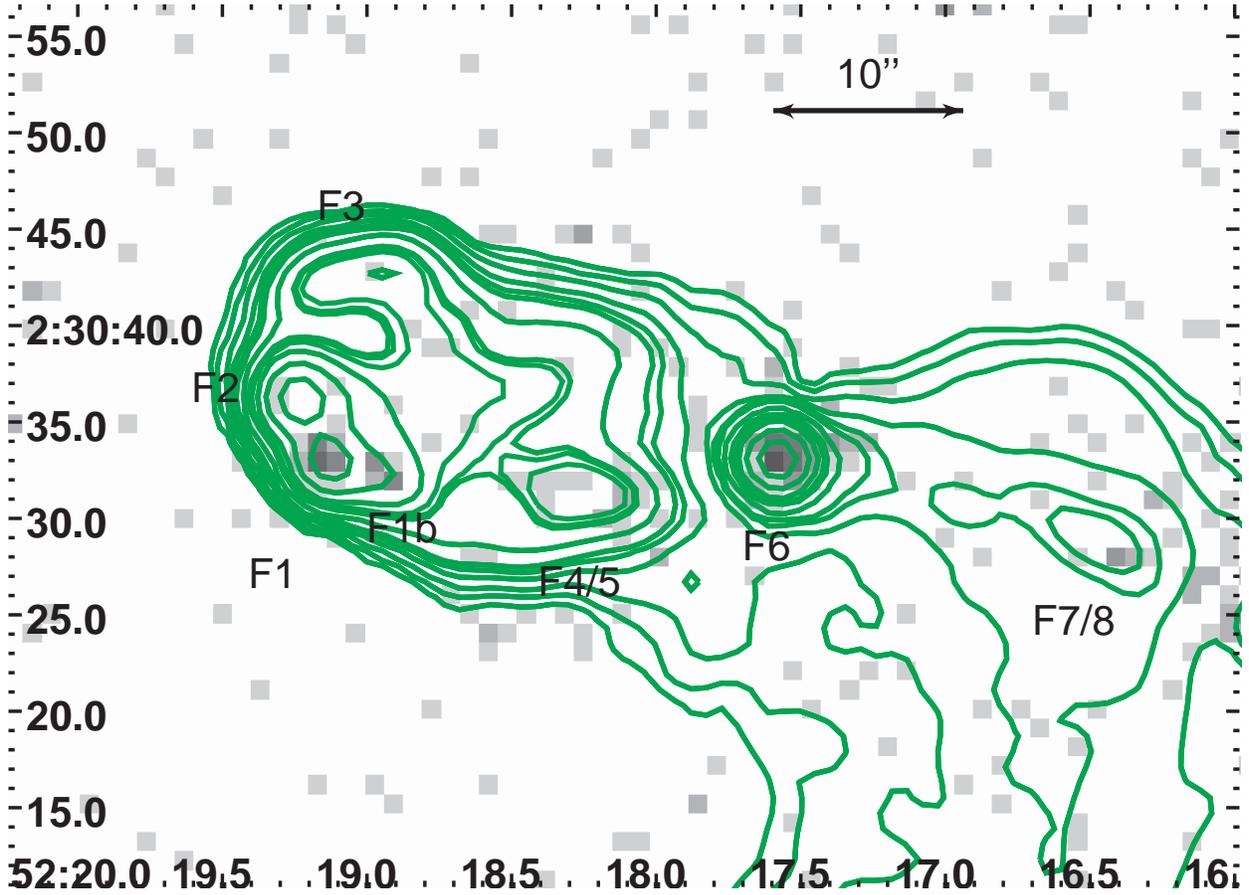}
\caption{Raw X-ray image (1 pixel=0.984$''$) in the 0.5-2.0 keV bandpass with 8.4 GHz radio contours
overlaid.  We clearly detect X-ray emission from radio knots F1, F4/5, and F6, and
the jet-like feature F7/8.  The
X-ray emission from knot F6 is clearly extended, whereas the radio emission is
point like.  We also detect a pointlike source of X-rays $\sim$4$''$ from radio knots F1
but closer to the nucleus.  There is no clear radio knot corresponding to this X-ray
knot.  We have labeled this feature F1b.  The radio contours correspond to flux densities
of 0.18, 0.35, 0.6, 0.9, 1.4, 2.0, 3.3, 4.4, 11.5, 22.0, and 35.0 mJy per beam.
The resolution of the radio data is 0.75$''$$\times$0.75$''$.}\label{elobeovl}
\end{figure}

As shown in Figure~\ref{elobeovl}, small-scale X-ray emission is
associated with multiple radio components in the eastern lobe: these
include the radio components F7/F8, F6, F5, F1 and a distinct radio component upstream of F1
which we denote F1b. There is no detected X-ray emission associated
with the hotspot components F2 or F3 in the E lobe or with the larger
hotspots P1 and P2 in the western lobe. Figure~\ref{wings} shows
that there is also large-scale diffuse emission associated with the
western lobe and with both the extended wings.
The position of the active nucleus in the X-ray data
using the nominal aspect solution agrees with the radio
position to better than 0.5$''$.  We have not attempted to improve the
aspect solution beyond this.

\subsection{Compact components}

We begin by considering the compact components in the E lobe shown
in Figure~\ref{elobeovl}. For the
X-ray emission associated with the radio hotspots F1 and F6, there are
enough counts to fit a power-law spectrum. We find best-fitting photon
indices of $1.75^{+0.4}_{-0.3}$ and $1.7_{-0.2}^{+0.3}$ for F1 and F6
respectively, and normalizations corresponding to flux densities at 1 keV of
$0.9 \pm 0.2$ and $2.3 \pm 0.2$ nJy. For the other three compact
components there are not enough counts to extract a spectrum, so
we estimated the flux densities based on the count rates assuming a
photon index of 2. (The flux density is insensitive to our assumed
photon index: for a photon index of 1.5 the flux densities would
decrease by $\sim 10$\%.) Flux densities for all the eastern lobe
components are tabulated in Table \ref{compact}.

We use the 8.4 GHz maps \citep{bla92} to measure radio flux
densities corresponding to the X-ray detections. For compact radio
components it is possible to extract radio flux densities either by
integrating over the whole X-ray extraction region (with local
background subtraction) or by fitting a model, such as a homogeneous
sphere (as was done by \citet{hard04}) or a Gaussian. In Table
\ref{compact} we tabulate radio flux densities extracted using both
methods, where they were both possible, which is true of the compact
components F1 and F6; this illustrates the range of
possible radio flux densities that can be said to correspond to the
X-ray emission. We also tabulate the ratio of X-ray to radio flux.

The compact components F1 and F6 are detected with HST; the HST image
with X-ray contours overlaid is shown in Figure~\ref{xknotovl}.
The radio-optical-X-ray spectra of these two knots are shown in Figure~\ref{knotspec}.
We have fitted synchrotron models to these data, and
both of these components are well fitted by a
single standard continuous injection model \citep{hea87}.
As pointed out by \citet{hard04}, inverse-Compton
models require extreme conditions to reproduce the observed X-ray flux
densities.  The knots must either be far from equipartition
or they must be moving relativistically at an angle close to
the line of sight.
A synchrotron model seems to be the only plausible
explanation for the X-ray emission from F1 and F6. For the other
components, the existing short HST observation does not give us
useful constraints (these features are resolved at 8.4 GHz with an
$0\farcs25$ beam, and therefore spread over many HST WFC pixels)
and we do not have X-ray spectral measurements. However, the radio to
X-ray ratios for the other detected components (Table \ref{compact})
are very similar to those measured for F1 and F6, so it would seem
plausible that they have the same emission mechanism.

An interesting feature of this interpretation is that some of the
X-ray detected components of 3C\,403 are morphologically part of the
jet rather than being hotspots: this is true of F1b, F5/4 and (the
clearest case) F7/8, and it is also debatable whether F6 should be
considered a jet knot rather than a primary hotspot. X-ray emission
from hotspots is of course seen in many other sources, and has been
attributed to synchrotron emission in some cases (e.g. 3C\,390.3,
\citet{har98}), although 3C\,403 is one of the
clearest cases of synchrotron rather than inverse-Compton emission
\citep{hard04}. However, 3C\,403 is one of the best examples of
synchrotron X-ray emission from the jet of a powerful {\it
narrow-line} radio galaxy.
In unified models of FRII radio galaxies, the jets of narrow-line radio
galaxies are at large angles to the line of sight \citep{bar89}; this
is true even for lower-luminosity objects, whose unification partners
must be broad-line radio galaxies rather than quasars \citep{hard98b}.
In both cases, the numbers of the different classes of
objects observed require that the narrow-line objects be at angles
greater than $45\degr$ to the line of sight. Such large angles
effectively rule out the possibility of applying the popular beamed
inverse-Compton model \citep{tav00} to the jet in
3C\,403. We have calculated the expected flux density in this model
for the jet component F7/8 for a range of angles and bulk Lorentz
factors, and find that the observed flux density can only be
reproduced with large bulk Lorentz factors ($\Gamma > 7$) and small
angles to the line of sight ($\theta < 8\degr$). While the large bulk
Lorentz factors cannot be ruled out, such small angles to the line of
sight are inconsistent with everything that is known about 3C\,403. A
synchrotron model appears to be the only viable one here.
It is unclear what fraction of FRII jets are dominated
in the X-ray band by synchrotron emission,
but they may be common.  The inverse-Compton process
may dominate only in those sources whose jets
are highly beamed.

\begin{figure}
\plotone{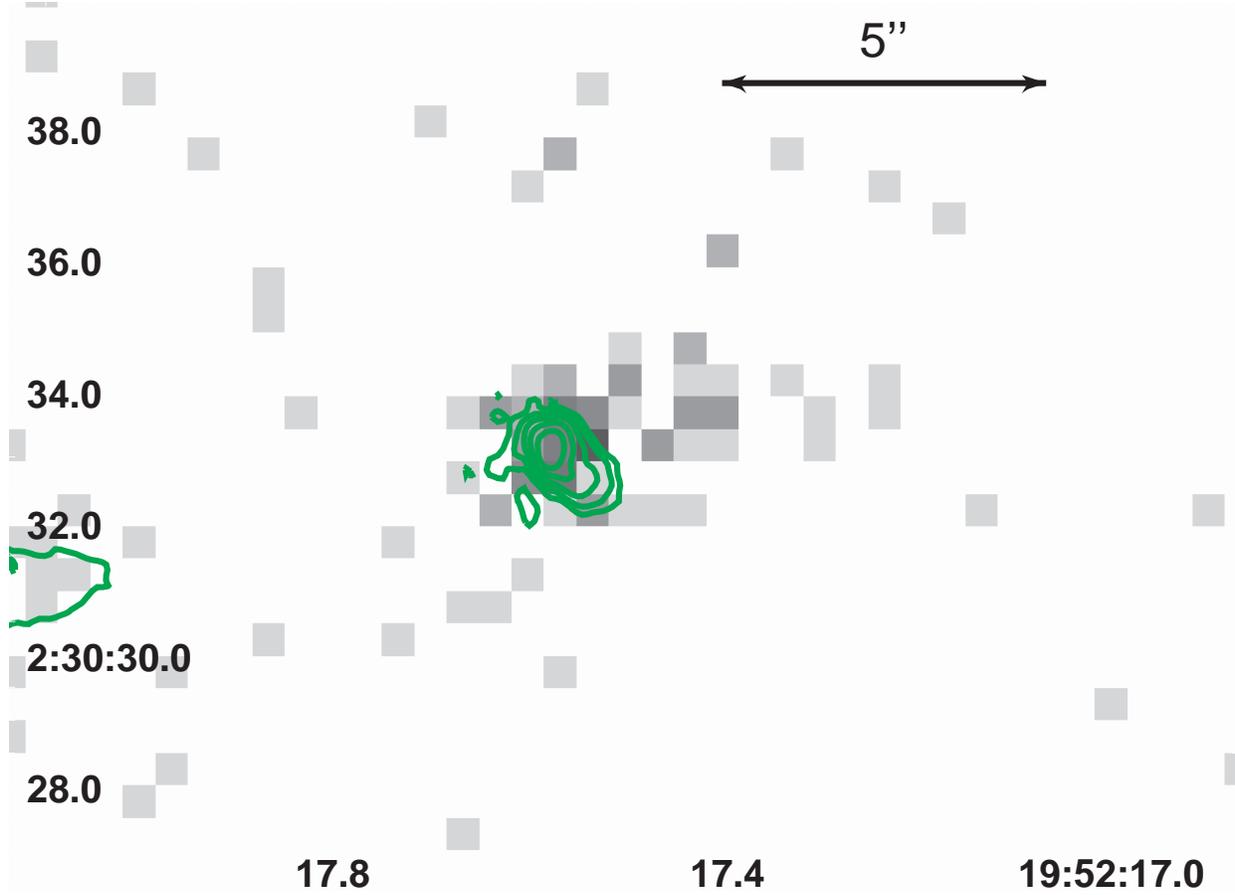}
\caption{Raw X-ray image (1 pixel=0.492$''$) of knot F6 in the 0.5-2.0 keV bandpass with 8.4 GHz radio contours
overlaid.  Note the X-ray extension to the W of the main peak, and the radio extension
to the SW.  The radio contours correspond to flux densities of 0.04, 0.2, 0.6,
2.0, and 4.0 mJy per beam.  The resolution of the radio data is 0.25$''$$\times$0.25$''$.}\label{f6knot}
\end{figure}

\begin{figure}
\plotone{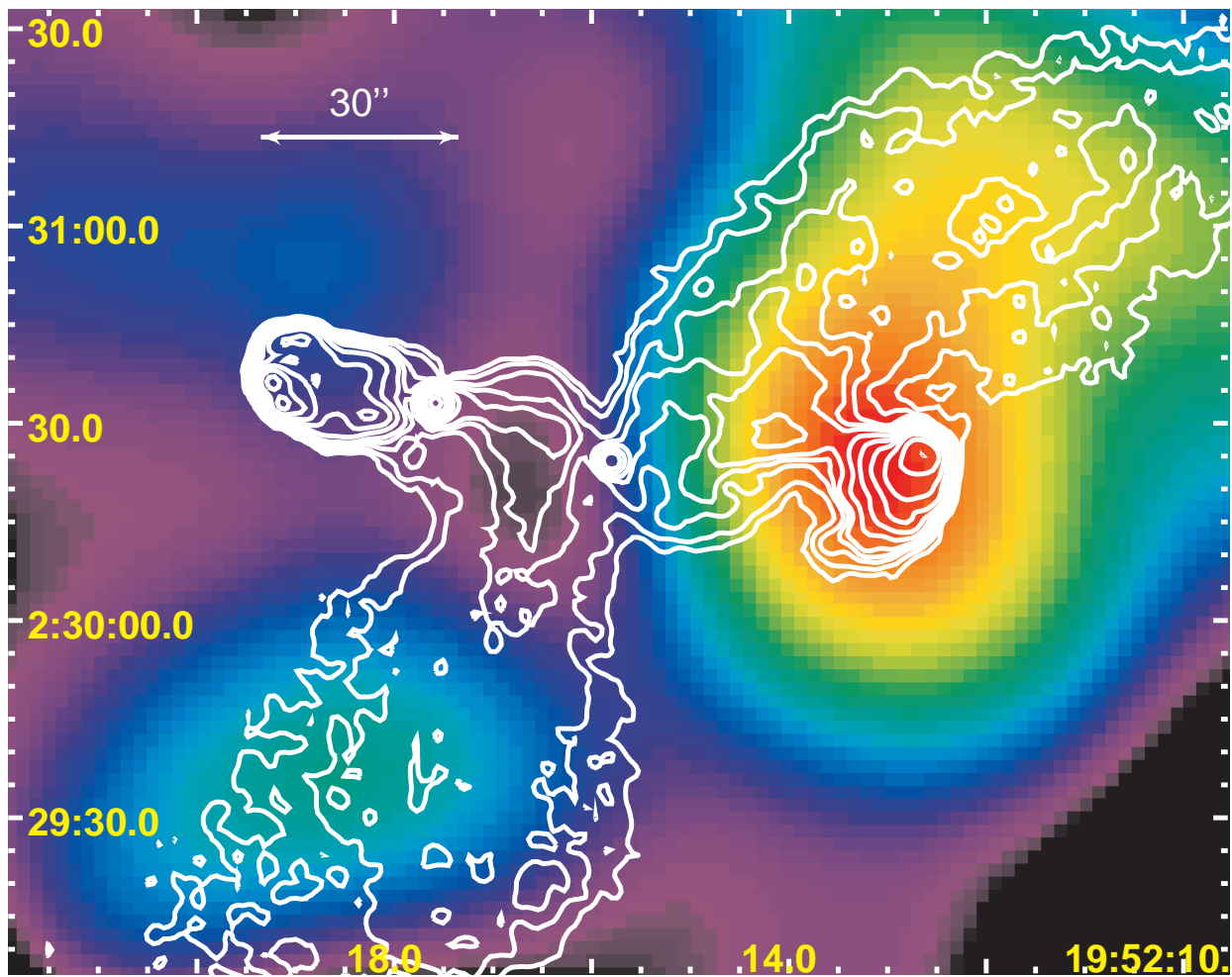}
\caption{Gaussian smoothed ($\sigma$=20$''$), exposure corrected, background
subtracted X-ray image in the 0.5-2.0 keV bandpass with all small scale features,
including the nucleus, the knots, and the central ISM, removed.  Radio
contours (8.4 GHz) are overlaid.  The low surface brightness diffuse X-ray
components associated with the wings and the western radio lobe are
visible.  The orange, green and blue regions
correspond to surface brightnesses of 3.2, 1.9, and 1.0$\times$10$^{-3}$ cts s$^{-1}$ arcmin$^{-2}$,
respectively.  The radio contours correspond to flux densities of 0.2, 0.4, 0.6, 1.1, 1.5, 2.5, 4.5, 12.0,
20.0, and 30.0 mJy per beam.  The resolution of the radio data is 2.5$''$$\times$2.5$''$.}\label{wings}
\end{figure}

\begin{figure}
\plotone{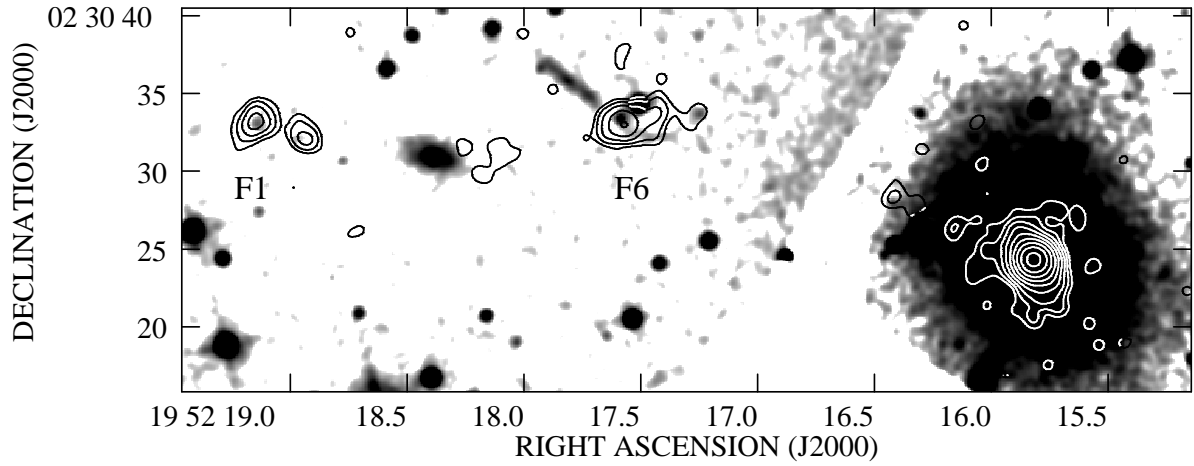}
\caption{Mosaic HST image of knots F1 and F6 with X-ray (0.5-5.0 keV band) contours overlaid.  The
host galaxy is seen to the west.  The X-ray data has been smoothed with a 0.5$''$
(FWHM) Gaussian.}\label{xknotovl}
\end{figure}

\begin{figure}
\plottwo{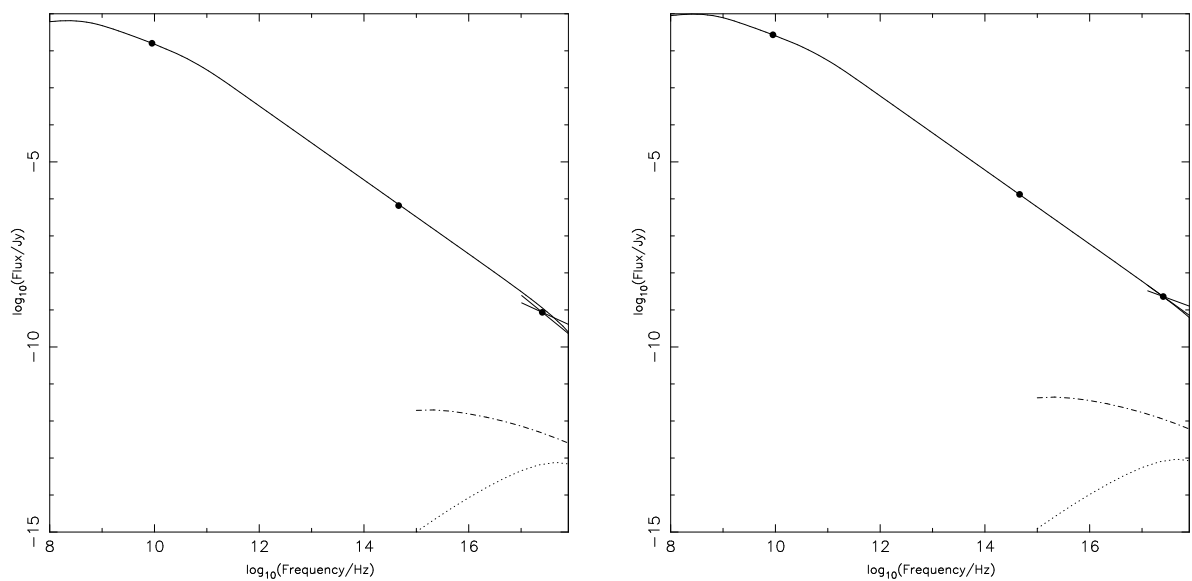}{f12b.ps}
\caption{The radio through X-ray spectrum of
knots F1 (left) and F6 (right), both fitted with a standard continuous
injection electron spectrum \citep{hea87} with an
electron energy break at an electron Lorentz factor $\gamma$ = 14000
and a minimum energy corresponding to $\gamma$ = 1000, and assuming an
equipartition magnetic field strength. The dot-dashed and dotted lines
show the contributions expected from synchrotron self-Compton emission
and inverse-Compton scattering of the CMB, respectively.}\label{knotspec}
\end{figure}

Finally, we wish to draw attention to two other interesting features
of the compact X-ray emission. The first is the non-detection of the
hotspot component F2. The upper limit on the X-ray emission from this
hotspot is an order of magnitude below the level at which it would
have been detected if its radio to X-ray ratio had been the same as
the detected components'. In a synchrotron interpretation, this
requires a difference in the particle acceleration properties of F1
and F2. It is possible that F2 is a relic hotspot that is now
detached from the jet flow and that has no ongoing particle
acceleration. The second interesting feature is the clear extension of
the brightest X-ray feature, F6, back towards the nucleus.  Figure~\ref{f6knot}
contains an X-ray image of the knot F6 in the 0.5-2.0 keV band with
8.4 GHz radio contours overlaid.  The peaks of the radio and
X-ray emission are well aligned, and both knots are clearly extended.
The X-ray extension is to the W of the main peak, and the radio
extension to the SW.  
There is no clear radio counterpart to the X-ray extension and if the extension is
considered to be radio-related, it has an X-ray to radio flux density
ratio more than a factor 10 above most of the other features in the
jets (based on an upper limit to the possible associated radio
emission). This sort of X-ray to radio ratio is not impossible for a
synchrotron feature --- knots in the Centaurus A jet show similar
properties --- but its location just upstream of a bright hotspot or
jet knot is certainly intriguing, and reminiscent of the `offsets'
seen in the jets of FRI radio galaxies \citep{kra02,hard03}.

\subsection{Extended components}

Extended X-ray emission from the lobes of FRII radio sources is normally
attributed to inverse-Compton emission 
(e.g. \citet{hard02,bru02,iso02,com03,bel04,cro04a}).
Typically the X-ray flux densities
measured from the lobes are close to, or slightly above, the value
expected for inverse-Compton scattering of the microwave background
radiation by lobes with magnetic field strengths close to the
equipartition value. The detected extended emission from the W lobe
and wings of 3C\,403 is too faint and diffuse for us to fit a
spectrum, so that we cannot test the inverse-Compton model's
prediction of a flat (photon index 1.5) X-ray spectrum. However, we
can estimate the 1-keV flux densities of the various components from
the observed count rate. Here we assume a photon index of 1.5 in
converting between counts and flux density. The results are tabulated
in Table \ref{extended}.
The X-ray counts are background subtracted, and the uncertainties are statistical
(Poisson).

We measured corresponding radio fluxes from 1.4 GHz and 8.4 GHz radio
maps (the low-resolution map of \citet{bla92} and a map made from the
1.4-GHz B-configuration data of \citet{den02}). These
are also tabulated in Table \ref{extended}. Modeling the lobes and
hotspots as being in the plane of the sky, we then calculated the
expected inverse-Compton flux density at equipartition. To do this we
assumed no relativistic protons, a low-energy electron energy cutoff 
corresponding to $\gamma = 10$,
and a low-energy electron energy index of 2 (corresponding to
$\alpha = 0.5$.  A high-energy break and cutoff were included to fit to
the steep spectral index between 1.4 and 8.4 GHz, and the radio spectra were
normalized to the appropriate fraction of the 178-MHz flux density of
28.3 Jy (see \citet{lea97}) --- this is justified given the very
uniform spectral index found by \citet{den02}. The
inverse-Compton calculation was then carried out using the code of
\citet{hard98a}.  We have assumed cylindrical geometry for the wings,
and spherical geometry for the W lobe.
The results are tabulated in Table~\ref{extended}.

The ratios of observed to predicted X-ray flux are around 4 for the
two wings, which is not particularly unusual for radio galaxies
\citep{cro04b} and suggests field strengths within a factor of $\sim
2$ of the equipartition value in the wings. The discrepancy with the
equipartition prediction would be further reduced if there were any
projection of the wings. If we assume no projection, and take the
X-ray flux density as giving a magnetic field strength measurement,
then the electron energy density dominates over the magnetic field
energy density in these lobes by around a factor 20, and the
non-thermal pressure in the wings is about $10^{-12}$ dyn cm$^{-2}$ (with large
uncertainties because of the large formal error on the count
measurement). The western lobe, on the other hand, has a ratio of
observed to predicted flux that is much larger than is typical, and
the distribution of X-ray counts in this region is also not what would
be expected from a simple inverse-Compton model, in which it should
more or less match the low-frequency radio emission (Fig.
\ref{wings}). The X-ray emission is not peaked at the radio peak
in the N end of the lobe, rather the X-ray emission uniformly fills the lobe.
Therefore, this emission cannot be explained by inverse-Compton scattering
of CMB photons by the electrons responsible for the radio emission.
Because of the differences in radio and X-ray morphology, this emission is
unlikely to be a shock heated shell of gas surrounding the lobe
that would be created if the lobe were expanding into the ISM supersonically
\citep{hei98,kra03}.
It is possible that the X-ray emission from the W lobe is due to
inverse-Compton scattering of beamed IR/optical photons from the active nucleus
by a lower energy, unobservable, population of relativistic electrons.  Such a model
has been used to explain diffuse X-ray emission from the western lobe
of the quasar 3C 207 \citep{bru97,bru02}.

\section{Discussion - The Nature of X-shaped Radio Galaxies}

As summarized in the introduction, there are two general scenarios commonly
invoked to explain the X-shaped morphology of radio galaxies such as 3C\,403.
One line of argument suggests that the X-shaped morphology is a hydrodynamic
phenomenon, and the other is that X-shaped radio galaxies are formed as the
result of a sudden change in the jet axis of AGN, perhaps as the result
of a galaxy merger and/or the merger of two supermassive black holes.
Within the hydrodynamic scenario, there are two general models in the
literature: the X-shaped
structure is the result of strong (perhaps highly supersonic) backflow
and subsequent buoyant evolution \citep{lea84,wor95}, or that both the lobes and wings are
highly supersonic (relative to the ambient ISM/ICM)
and the observed structures are the natural result of over-pressurized expansion/inflation
of the lobes into an asymmetric medium \citep{cap02}.
We prefer the backflow/buoyancy model because, in our view, the observed structures
of X-shaped radio galaxies
are the natural consequences of strong backflows in an asymmetric medium,
and there is no indication that 3C\,403's lobes are highly overpressured.

We know that strong backflows exist in radio galaxies, and that there are many
cases where the morphology of the ambient medium in clusters, groups, and isolated
ellipticals is complex.
The dynamics of the backflow of material from the jet
head are determined by several factors including the
jet velocity, the density contrast between the jet and the ambient medium, and
the temperature and density profile of the medium into which the jet is propagating
\citep{lea84,hardee92}.  
In particular, lighter jets (i.e. jets with a larger values of $\rho _{ISM}$/$\rho _{jet}$)
have more powerful backflows than heavier jets.
Backflow velocities ranging from several thousand km s$^{-1}$ to
0.1$c$ have been estimated for a sample of double lobed 3CR radio galaxies \citep{alex87}.
Thus backflow velocities can be orders of magnitude
higher than the advance speed of the jet head.  Once the backflow material reaches
a strong pressure gradient, it will rapidly decelerate, bend (depending on the details
of the gradient), and subsequently it will
be driven primarily by buoyancy forces (i.e. subsonically).
The density and pressure morphologies of the hot gas in clusters, groups, and
elliptical galaxies is complex.  If {\em Chandra} observations of clusters of
galaxies are any guide, there are no such things as spherically symmetric gas halos
sitting idly in their gravitating dark matter potentials (Markevitch, 2004, private communication).
In our view, there is no compelling reason to invoke more complex phenomena such
as merging supermassive black holes to explain X-shaped radio galaxies.

The arguments against the hydrodynamic models are based primarily on
three considerations: the radio morphology of these objects, evidence for recent
mergers, and radio spectral differences of the wings and lobes.
None of these are strong arguments in our view and we suggest
that it is quite plausible (and even natural) for backflow/buoyancy to create the structures that
are observed.  
The most important of the three are those based on the radio morphology.
In some sources, the wings (in projection) are longer than the lobes
(as is the case in 3C\,403) \citep{den02}, and in
others the alignment of the wings and lobes relative to the nucleus is
sometimes offset (e.g. NGC 326) \citep{gop03}.  Qualitative arguments are often made that the
observed morphologies are unlikely to be generally formed by hydrodynamic flows.
The implicit assumption in these arguments is that the gas atmosphere that the jet
is propagating into is roughly spherically symmetric.
The observed radio structures will depend critically on the distribution of the
gas into which the jet is propagating.
Similarly, the general paradigm suggests that the lobes of FR IIs are advancing supersonically into
the ambient medium, and 
it has been argued that it is not plausible that the wings and lobes are approximately
the same length if the lobes are evolving supersonically and the wings buoyantly \citep{lea91}.
This claim is true \emph{if} the lobes are advancing supersonically into the ambient medium, but there
is no direct evidence to support this claim.
In all cases where the internal and external pressures of FR II lobes have been measured to date,
the lobes appear to be close to pressure equilibrium with their surroundings, as we 
observe in 3C\,403 \citep{hard02,bel04,cro04a}, implying relatively slow expansion.
The complex radio morphology of 3C\,403 is clearly different from the canonical
lobe/hotspot morphology of classic FR II radio galaxies such as Cyg A,
and its radio luminosity lies on the FRI/FRII dividing line.  Both of these suggest
that 3C\,403 is among the least powerful sources in this class.
The inner radio lobes of M87, a radio galaxy on the FRI/II luminosity boundary,
have detached from the shock propagating through the ICM of the Virgo cluster
and are thus evolving buoyantly/subsonically \citep{for04}.

Additionally, some have argued that there is evidence for recent mergers in the host galaxies of
some of these sources.  This is certainly not a general
conclusion as 3C\,403, one of the
best studied X-shaped radio galaxies, resides in a relatively poor environment
and shows little evidence for a recent (within the last $\sim$10$^8$ years) merger.
HST observations of the host galaxy of 3C 403 indicate the presence of weak dust lanes, suggestive
of a much older merger, certainly older than the age of the radio source, estimated to be
$\sim$16 Myr based on radio spectral analysis \citep{den02}.
Finally, multispectral radio observations of several winged sources (e.g. NGC 326 \citep{eke78}
and B2 0828+32 \citep{kle95}) suggest that the plasma in the wings
is considerably older than that in the lobes.
Variations in radio spectral index across the wings and lobes would be a natural result
if the jet axis were slowly precessing.  However,
this observational result is also consistent with the hydrodynamic backflow models in our view
for two reasons.  First, the plasma in the wings is naturally older than that in the lobes
since it traveled back from the jet head to its present location in the wings.
Second, the evolutionary history of the material in the wings is clearly different from that of the
material currently in the lobes in the backflow model.  The spectral differences could well be
telling us something about the hydrodynamics and the nature of shock acceleration
of the backflow, and be unrelated to radiative energy loss of the most
energetic particles.  It is impossible to draw definitive conclusions
from spectral differences between the wings and lobes.  

The quality of our data presented in this paper is not sufficient for us to put strong
constraints on the dynamics of the lobes/wings interaction with the ISM
and make a definitive, general statement regarding the nature
of X-shaped radio galaxies.  We have detected non-axisymmetric structures in the
central region of the host galaxy.  
This result is consistent with the various hydrodynamic scenarios that have
been invoked to explain this phenomenon.
Deeper observations of this object and others
with {\em Chandra} and/or XMM/Newton are required
to better understand the interaction between the gaseous corona
and the radio lobes in order to resolve this issue.

\section{Conclusions}

The primary results of this work include the following:

\begin{itemize}

\item We have detected X-ray emission from the hot ISM, the active nucleus,
several radio knots, and diffuse emission from the lobes and wings of
the nearby X-shaped radio galaxy 3C\,403.

\item Within $\sim$5 kpc of the nucleus, the morphology of the hot gas is highly ($\epsilon\sim$0.57) elliptic.
This is consistent with the hypothesis that the X-shaped radio morphology is the result
of the propagation of jets in non-axisymmetric atmospheres.
The distribution of the stars in the central 5 kpc is considerably
less elliptical ($\epsilon\sim$0.2).
Our data are not of sufficient quality for us to make a good measurement of
the density/pressure profile of the
hot gas on larger scales, but within the uncertainties, the lobes and wings are in approximate 
pressure equilibrium with the hot corona.

\item We have detected X-ray and optical emission from radio knots and hotspots.
Based on their radio/optical/X-ray, we attribute the emission in all cases to
synchrotron emission, not IC scattering from the CMB.
All lines of evidence suggest that the jet and radio lobes lie at a large angle to
the line of sight, so the effects of relativistic beaming are negligible.
This is one of the best examples to date of synchrotron X-ray emission from the compact radio
features of a narrow-line FR II radio galaxy.

\item The X-ray spectrum of the nucleus is well modeled by two power laws plus
a broad, fluorescent Fe line from cold material.  The first power law is
heavily absorbed and is buried deeply in the torus.  The second power law is
less heavily absorbed, and we attribute this to X-ray emission from an unresolved
jet.  The ratio of X-ray to radio flux is consistent with that measured for
the nuclei of other radio galaxies.

\item Diffuse X-ray emission is detected from the lobes and wings.  This emission
can generally be explained as IC scattering from the CMB if we assume moderate
departure from equipartition.

\end{itemize}

\acknowledgements

This work was supported by NASA contracts NAS8-38248, NAS8-39073, the
Royal Society, the Chandra X-ray Center, and the Smithsonian Institution.
We would also like to thank the anonymous referee for comments and
suggestions that improved this paper.

\clearpage

\clearpage

\clearpage

\clearpage

\clearpage

\clearpage

\clearpage

\clearpage

\clearpage

\clearpage

\clearpage

\clearpage

\clearpage

\begin{table}
\begin{center}
\begin{tabular}{|c|c|c|}\hline
Count rate &          & 4.71$\times$10$^{-2}$ cts s$^{-1}$ \\ \hline\hline
APEC       & $k_BT$   & 0.3$\pm$0.04 keV \\ \hline
           & $Z$      & 1.0$\times$Solar (fixed) \\ \hline
           & $N_H$    & 1.54$\times$10$^{21}$ cm$^{-2}$ (fixed - Galactic) \\ \hline
Flux       &          & 3.6$\times$10$^{-14}$ ergs cm$^{-2}$ s$^{-1}$ \\ \hline
Luminosity & $L_X$    & 2.9$\times$10$^{41}$ ergs s$^{-1}$ \\ \hline\hline
First PL   & Index    & 1.70$\pm$0.15 \\ \hline
           & $N_H$    & 4.0$\pm$0.2$\times$10$^{23}$ cm$^{-2}$ \\ \hline
Flux       &          & 1.31$\times$10$^{-11}$ ergs cm$^{-2}$ s$^{-1}$ \\ \hline
Luminosity & $L_X$    & 1.07$\times$10$^{44}$ ergs s$^{-1}$ \\ \hline\hline
Second PL  & Index    & 2.0 (fixed)  \\ \hline
           & $N_H$    & 4.0$\times$10$^{21}$ cm$^{-2}$ (fixed) \\ \hline
Flux       &          & 8.75$\times$10$^{-14}$ ergs cm$^{-2}$ s$^{-1}$ \\ \hline
Luminosity & $L_X$    & 7.11$\times$10$^{41}$ ergs s$^{-1}$ \\ \hline\hline
Fe line    & Centroid & 6.31$\pm$0.04 keV \\ \hline
           & Width (r.m.s.)  & 80$\pm$50 eV \\ \hline
           & EW       & 244$\pm$20 eV \\ \hline\hline
\end{tabular}
\caption{Summary of spectral parameters for nucleus and hot ISM.  All uncertainties
are at 90\% confidence.  The count rate is in the 0.25-10.0 keV
band.  The fluxes and luminosities listed for each component are unabsorbed in the
0.25-10.0 keV band.  The equivalent width of the Fe line is relative to the
first power-law component.}\label{specparms}
\end{center}
\end{table}

\clearpage

\begin{table}
\caption{X-ray (1 keV) and radio flux densities (8.4 GHz) for compact components in the
  E~lobe.  The radio to X-ray photon index was assumed to be 2 unless otherwise stated.}
\label{compact}
\begin{tabular}{|l|r|r|r|r|r|r|}
\hline
Name &0.5-5.0 keV & X-ray flux density & Photon    & Radio flux    & Radio flux   & X-ray/radio\\
     &counts      & (nJy at 1 keV) & index  & (model) (mJy) & (integ.) (mJy)      & ratio ($\times 10^6$)\\
\hline
F1&$34 \pm 6$&$0.9 \pm 0.2$&$1.75^{+0.4}_{-0.3}$&16&25&0.06--0.04\\
F1b&$15 \pm 4$&$0.5 \pm 0.1$&2&&7.7&0.06\\
F2&$<4$&$<0.13$&2&&27&$<0.004$\\
F5/4&$14 \pm 4$&$0.5 \pm 0.1$&2&&10.8&0.05\\
F6&$83 \pm 9$&$2.3 \pm 0.2$&$1.7_{-0.2}^{+0.3}$&27&41&0.09--0.06\\
F7/8&$9 \pm 3$&$0.3 \pm 0.1$&2&&2.6&0.12\\
\hline
\end{tabular}
\end{table}

\clearpage

\begin{table}
{\small
\caption{Flux densities and predictions for the extended components of 3C\,403\tablenotemark{a}\tablenotetext{a}{The cylinder and sphere dimensions are the length$\times$radius and radius, respectively.}.}
\label{extended}
\begin{tabular}{|l|c|c|c|c|c|c|c|}
\hline
Name   & 0.5-5.0 keV& 1-keV flux & 1.4-GHz  & 8.4-GHz & Geometry & Predicted (1 keV) & Ratio \\
       & counts     & (nJy)      & flux (Jy)& flux (Jy) &          & IC (nJy)          & (Obs/Pred)\\ \hline
N wing & $63 \pm 29$& 1.6        & 0.78     & 0.27    & Cylinder, & 0.35              & 4.6\\
       &            &            &          &         &$90'' \times 24''$ &           & \\
S wing & $53 \pm 28$& 1.4        & 0.71     & 0.24    & Cylinder, &0.34               &4.0 \\
       &            &            &          &         &$90'' \times 25''$ &           & \\
W lobe &$55 \pm 14$ & 1.4        & 1.88     & 0.50    & Sphere,           & 0.1       & 14\\
       &            &            &          &         & 7\farcs5          &           & \\ \hline
\end{tabular}
}
\end{table}

\end{document}